\documentclass[a4paper,11pt]{article}
\pdfoutput=1 

\usepackage{jheppub}                                         
\usepackage{color}
\usepackage{tensor}
\usepackage[usenames,dvipsnames]{xcolor}

\newcommand{\be}{\begin{equation}}
\newcommand{\ee}{\end{equation}}
\newcommand{\ba}{\begin{eqnarray}}
\newcommand{\ea}{\end{eqnarray}}

\newcommand{\beq}{\begin{equation}}
\newcommand{\eeq}{\end{equation}}
\newcommand{\beqa}{\begin{eqnarray}}
\newcommand{\eeqa}{\end{eqnarray}}
\newcommand{\nn}{\nonumber}

\title{\boldmath Thermalon mediated phase transitions in Gauss-Bonnet gravity}

\author[a]{Robie A. Hennigar,}
\author[a]{Robert B. Mann,}
\author[a]{and Saoussen Mbarek}

\affiliation[a]{Department of Physics and Astronomy, University of Waterloo, \\
Waterloo, Ontario, Canada, N2L 3G1, Canada}

\emailAdd{rhennigar@uwaterloo.ca}
\emailAdd{rbmann@sciborg.uwaterloo.ca}
\emailAdd{smbarek@uwaterloo.ca}

\abstract{Thermalons can mediate phase transitions between different vacua in higher curvature gravity, potentially changing the asymptotic structure of the spacetime.  Treating the cosmological constant as  a dynamical parameter, we study these phase transitions in the context of extended thermodynamic phase space.  We find that in addition to the AdS to dS phase transitions previously studied, thermal AdS space can undergo a phase transition to an asymptotically flat black hole geometry.  In the context of AdS to AdS transitions, we comment on the similarities and differences between thermalon transitions and the Hawking-Page transition. }
 
\begin{document} 
\maketitle
\flushbottom

\section{Introduction}

Phase transitions in gravitational physics have been a subject of interest for the last few decades. 
More than 35 years ago, Coleman and de Luccia discussed {\it gravitational instantons}, showing that coupling a scalar field to a dynamical metric can lead to phase transitions between two competing vacua with different cosmological constants \cite{Coleman:1977py,Coleman:1980aw}.  These transitions proceed via the nucleation of expanding bubbles of true vacuum within the false vacuum when the free energy of the true vacuum becomes smaller than that of the false vacuum.   
Mechanisms of this kind have been utilized in various proposed solutions to the cosmological constant problem \cite{Brown:1987dd, Brown:1988kg}.  Another classic example  of a gravitational phase transition is the {\it Hawking-Page transition} \cite{Hawking:1982dh}, which has significance in various proposed gauge/gravity dualities. This phenomenon is a first order phase transition between thermal Anti de Sitter (AdS) space and the Schwarzschild-AdS black hole, with the latter becoming thermodynamically preferred (i.e. lower in free energy) above a certain critical temperature. 

A number of recent studies have focused on {\it thermalon mediated phase transitions} in higher curvature gravity \cite{Camanho:2013uda, Camanho:2015zqa,  Camanho:2012da}.  These phase transitions  proceed via the nucleation of spherical shells, called {\it thermalons}, that separate spacetime into two regions described by different branches of the solution, hosting a black hole in the interior. Above a certain critical temperature  the thermalon configuration is thermodynamically preferred to finite temperature AdS space . The thermalon, once formed, is dynamically unstable and expands to fill all space in finite time, effectively changing the asymptotic structure of the spacetime.  In a study focusing on a fixed value of the cosmological constant, it has been shown 
\cite{Camanho:2015zqa, Nojiri:2001pm} that thermal AdS space can undergo a thermalon-mediated phase transition to an asymptotically dS black hole geometry---in some sense a generalized version of the Hawking-Page transition. 

In the present work, we will be studying  thermalon mediated phase transitions in the context of extended phase space thermodynamics. In this framework, one promotes the cosmological constant to a thermodynamic variable, equivalent to   pressure in the first law \cite{Kastor:2009wy, Dolan:2010ha,Dolan:2011xt,Dolan:2011jm,Dolan:2012jh,Cvetic:2010jb,Larranaga:2011wd,Larranaga:2012pq,Kubiznak:2012wp,Gunasekaran:2012dq,Belhaj:2012bg,Lu:2012xu,Smailagic:2012cu,Hendi:2012um}. In general, the relationship between the cosmological constant and the thermodynamic pressure is
\beq\label{pressure_def}
P = - \frac{\Lambda}{8\pi G} = - 2 \Lambda
\eeq
where the last equality follows since we employ the normalization $16 \pi G = 1$ for consistency with ref.\cite{Camanho:2013uda}.
The corresponding conjugate quantity is the thermodynamic volume which is defined to ensure the validity of the extended first law 
\be 
\delta M = T \delta S + \sum_i \Omega_i \delta J_i + \Phi \delta Q + V \delta P \, ,
\ee
a result which follows from geometric arguments \cite{Kastor:2009wy}, and which renders
the Smarr relation consistent  with Eulerian scaling. A natural consequence of the extended phase space paradigm is that it allows us to understand   mass as the gravitational analogue of the enthalpy of a black hole rather than the total energy of the system, which has far-reaching consequences.  Numerous studies have found examples of van der Waals behaviour, (mutiple)-reentrant phase transitions,  tricritical points (analogous to the triple point of water), and isolated critical points for a variety of asymptotically AdS black holes \cite{Kubiznak:2012wp, Gunasekaran:2012dq, Altamirano:2013ane, Altamirano:2013uqa, Hennigar:2015esa,Hennigar:2014cfa, Hennigar:2015wxa}.  These results demonstrate a deep connection between the physics of AdS black holes and the physics of simple thermal systems that has come to be called {\it black hole chemistry} \cite{Kubiznak:2014zwa}. Our motivation for using this framework comes from  the fact that it is particularly well suited for an exhaustive study of the thermodynamic phase space.  As a result, we will be able to explore the properties of these phase transitions as the pressure varies.
 
Our paper is organized as follows.  In the next section  we briefly review the basics of Lovelock gravity and the essentials of the thermalon mechanism.  In section 3 we specialize to the case of Gauss-Bonnet gravity where we study the stability, extended phase space thermodynamics, and phase structure of the thermalons.  When considering the phase behaviour of these systems, we employ the extended thermodynamic phase space formalism to exhaustively study how these transitions depend on the pressure (cosmological constant).  In the context of AdS $\to$ dS + black hole thermalon mediated phase transitions we recover the results of \cite{Camanho:2015zqa}.  Furthermore, by analyzing the behaviour of the free energy near the Nariai limit, we find that for a fixed value of the Gauss-Bonnet coupling, there is a minimum pressure below which thermalon mediated phase transitions are not possible.  We find that in the case where the pressure is vanishing, a phase transition between thermal AdS space and an asymptotically flat geometry with a black hole is possible for any range of temperature.  In the last section we comment on the similarities and differences between the thermalon mediated phase transition and the Hawking-Page transition in the regime of positive pressures.

\section{Lovelock gravity}

It is generally expected that in any attempt to perturbatively quantize gravity one will find that the standard Einstein-Hilbert action is modified by the addition of higher curvature terms.  Natural candidates for the higher curvature corrections are provided by Lovelock gravities, which are the unique theories that give rise to generally covariant field equations containing at most second order derivatives of the metric \cite{Lovelock:1971yv}.  

The action for Lovelock gravity can be written in terms of bulk and boundary terms as \cite{Camanho:2013uda},
\be 
{\cal I} = \sum_{k=0}^{k_{\rm max}} \frac{c_k}{d-2k} \left(\int_{\cal M} {\cal L}_k - \int_{\partial {\cal M}} {\cal Q}_k \right)
\ee
where the $c_k$ are coupling constants,  
\be 
{\cal L}_k = \epsilon_{a_1 \cdots a_d} R^{a_1a_2} \wedge \cdots \wedge R^{a_{2k-1} a_{2k}} \wedge e^{a_{2k+1}} \wedge \cdots \wedge e^{a_d},
\ee
and
\be 
{\cal Q}_k = k \int_0^1 d\xi \epsilon_{a_1 \cdots a_d} \theta^{a_1a_2} \wedge \mathfrak{F}^{a_3a_4} \wedge \cdots \wedge \mathfrak{F}^{a_{2k-1} a_{2k}} \wedge e^{a_{2k+1}} \wedge \cdots \wedge e^{a_d} \, .
\ee
Here $e^a = \tensor{e}{^a_\mu} dx^\mu$ is the vielbein 1-form, $R^{ab} = d\omega^{ab} + \tensor{\omega}{^a _c}\wedge \tensor{\omega}{^c ^b}$ is the curvature 2-form ($\omega^{ab}$ being the torsionless Levi-Civita spin connection), and $\mathfrak{F}^{ab} = R^{ab} + (\xi^2 - 1) \tensor{\theta}{^a _c} \wedge \tensor{\theta}{^c ^b}$ with $\theta^{ab}$ being the second fundamental form related to the extrinsic curvature via $\theta^{ab} = (n^a \tensor{K}{^b _c}  - n^b \tensor{K}{^b _c}) e^c$.

One can obtain spherically symmetric solutions to the field equations of this theory of the form,
\be\label{metricansatz} 
	ds^2 = -f(r) dt^2 + \frac{dr^2}{f(r)} + r^2 d\Omega^2_{(\sigma) d-2},
\ee
with  $d\Omega^2_{(\sigma) d-2} $ denoting the line element on a $(d-2)$-dimensional compact space of constant curvature ($\sigma = 1,0,-1$ denoting spherical, flat and hyperbolic topologies, respectively).  Making use of the notation $g = (\sigma - f) / r^2 $, the field equations are solved provided
\be\label{charpoly} 
	\Upsilon[g] = \sum_{k=0}^K c_kg^k =  \frac{M}{r^{d-1}}
\ee
where $M$ is a constant of integration identified as the mass parameter of the solution.  Note that here we have suppressed, for convenience, a factor proportional to the volume of the unit radius manifold whose metric is given by $d\Omega^2_{(\sigma) d-2} $.\footnote{Our mass is related to the standard definition via $\tilde{M} = \frac{16 \pi G }{(d-2) \Sigma^{(\sigma)}_{d-2}} M$. }

This paper is concerned with {\it thermalon-mediated phase transitions}.  These transitions proceed via the production of a thermodynamically favoured but dynamically unstable spherical shell, called the thermalon, which divides spacetime into two regions.  The case of interest is when the spacetime metric is continuous but not differentiable at the  junction, a condition which, in Einstein gravity, would require the junction shell to possess stress-energy but does not have such a requirement in higher curvature gravity---one can think of the higher curvature terms themselves as providing the matter source.  Since the thermalon is dynamically unstable, once formed it expands rapidly, reaching spatial infinity in finite time and therefore changing the asymptotic structure of the spacetime.  In this way, the thermalon can be considered to mediate a phase transition between two vacua with different asymptotic structure.  

We now turn our attention to a brief recapitulation of the junction conditions and thermalon properties as discussed in \cite{Camanho:2013uda}.  Here we are interested in the case where a timelike junction surface separates an inner region and an outer region, which we denote with a $``-"$ and $``+"$, respectively.  In particular, we will be interested in the scenario where the metric function describing the inner geometry, $f_-(r)$, is different from the metric function describing the outer geometry, $f_+(r)$.  To this end, we decompose the spacetime manifold: $\mathcal{M} = \mathcal{M}_- \cup (\Sigma \times \xi) \cup \mathcal{M}_+$ where $\Sigma$ is the junction hypersurface and $\xi \in [0,1]$ is a real parameter used to interpolate both regions.  

Since the thermalon is a finite temperature instanton, we take the Euclidean metric to be
\be\label{thermalonansatz} 
	ds^2 = f_\pm(r) dt^2 + \frac{dr^2}{f_\pm(r)} + r^2 d\Omega^2_{(\sigma) d-2} \, ,
\ee
and describe the junction with the parametric equations
\be 
	r = a(\tau), \quad t_\pm = T_\pm(\tau)
\ee
and induced metric
\be\label{hypersurfacemetric} 
	ds^2 = d\tau^2 + a(\tau)^2 d\Omega^2_{(\sigma) d-2} \, .
\ee
Note that writing the hypersurface metric in the form of \eqref{hypersurfacemetric} assumes that the condition
\be 
	f_\pm \dot{T}_\pm + \frac{\dot{a}^2}{f_\pm(a)} = 1
\ee
is satisfied for all $\tau$ (a dot representing a $\tau$ derivative).  In the case of the thermalon, which is characterized by the static configuration $\dot{a} = \ddot{a} = 0$, i.e.  $a(\tau) = a_\star$, this condition amounts to the physical statement that the temperature of the bubble is the same as seen from both sides,
\be\label{thermal_matching} 
	\sqrt{f_-(a_\star)} \beta_- = \sqrt{f_+(a_\star)} \beta_+ = \beta_0 
\ee
where $\beta_-$ is the inverse Hawking temperature of the inner black hole and $\beta_+$ is the inverse temperature seen by an observer at infinity.  

As discussed in detail in \cite{Camanho:2013uda}, the junction conditions for this set up (without matter) amount to the continuity of the cannonical momenta across the hypersurface $\Sigma$,
\be\label{cannonical_momentum} 
	\pi^+_{ab} = \pi^-_{ab}
\ee
where the cannonical momenta are computed via the variation of the boundary terms at the junction surface \cite{Camanho:2015ysa}
\be 
\delta {\cal I}_\partial = - \int_{\partial {\cal M}} d^{d-1} x \pi^{ab} \delta h_{ab} \, .
\ee
However, for the case just described, the cannonical momenta have only diagonal components, which are themselves all related by the constraint
\be\label{bianchi-identity} 
	\frac{d}{d\tau} \left(a^{d-2} \pi^\pm_{\tau\tau} \right) = (d-2) a^2\dot{a}\pi^\pm_{\varphi_i \varphi_i}
\ee
where $\varphi_i$ represent the angular coordinates on $\Sigma$.  Due to this Bianchi identity, only the $\tau\tau$ component of the canonical momenta matters.  In \cite{Camanho:2013uda} the calculation of $\pi^\pm_{\tau\tau}$ is discussed, and we repeat the result here:
\be 
	\Pi^\pm = \pi^\pm_{\tau\tau} = \frac{\sqrt{\dot{a}^2 + f_\pm(a)}}{a} \int_0^1 d\xi \, \Upsilon' \left[\frac{\sigma - \xi^2 f_\pm(a) + (1-\xi^2)\dot{a}^2}{a^2} \right] 
\ee
where the prime on $\Upsilon$ refers to the derivative with respect to its argument.
In terms of the quantity $\tilde{\Pi} = \Pi^+ - \Pi^-$ Eqs. \eqref{cannonical_momentum} and \eqref{bianchi-identity} read,
\be 
\tilde{\Pi} = \frac{d\tilde{\Pi}}{d\tau} = 0 \,
\ee
and we have for $\tilde{\Pi}$ the convenient shorthand
\be\label{pitildeexpr} 
\tilde{\Pi} = \int_{\sqrt{H-g_-}}^{\sqrt{H-g_+}} dx \, \Upsilon' [H - x^2]
\ee
where $H = (\sigma + \dot{a}^2)/a^2$. 
In the following section we shall specialize to the case of Gauss-Bonnet (GB) gravity.

\section{Gauss-Bonnet case}

Gauss-Bonnet gravity is the simplest extension of the Einstein-Hilbert action to include higher curvature Lovelock terms.  In the following we shall adopt for the normalization of the Lovelock couplings 
\be
	c_0 = \frac{-2\Lambda_d}{(d-1)(d-2)} = -2 \Lambda \, , \quad c_1 = 1 \, , \quad c_2 = \lambda . 
\ee 
Note that here our definitions differ from those in refs. \cite{Camanho:2013uda, Camanho:2015zqa,  Camanho:2012da} in two ways.  First, we have not assumed a particular sign for the cosmological term and written it in terms of the radius of curvature, $L$.  This decision is simply for convenience when we will later identify the cosmological constant as a pressure.  Note also here our introduction of the terminology ``$\Lambda_d$" where the dimension-dependent factors have been absorbed to make a more convenient shorthand.  Secondly, we have not rescaled the GB coupling by a power of $\Lambda$ to make it dimensionless, since doing so would introduce extra and unnecessary factors of the pressure, thereby complicating the analysis.  

The characteristic polynomial \eqref{charpoly} now reads
\be\label{GB_char_poly} 
	\Upsilon[g_\pm] = -2\Lambda + g_\pm + \lambda g_\pm^2 = \frac{M_\pm}{r^{d-1}} \, .
\ee 
Explicitly solving this for $g_\pm(r)$ yields
\be 
	g_\pm(r) = -\frac{1}{2 \lambda} \left[1 \pm \sqrt{1 +4\lambda\left(2\Lambda + \frac{M_\pm}{r^{d-1}}\right)  } \right] \, 
\ee
and so
\be 
f_\pm(r) = \sigma +  \frac{r^2}{2 \lambda} \left[1 \pm \sqrt{1 +4\lambda\left(2\Lambda + \frac{M_\pm}{r^{d-1}}\right)  } \right] \, .
\ee
A point of particular interest is that for each branch of the solution there is an \emph{effective cosmological constant} given by
\be\label{effectivecosmo} 
	\Lambda_\pm^{\textrm{eff}} = - \frac{1 \pm \sqrt{1+8\lambda\Lambda}}{2 \lambda} \, ,
\ee
and so the two branches describe two asymptotically distinct solutions. 
The effective cosmological constants are generally different, being equal only when $\lambda= -1/(8 \Lambda)$, a case that corresponds to Chern-Simons theory \cite{Crisostomo:2000bb}.

Expressing the junction condition $\tilde{\Pi} = 0$ in the more convenient form $\dot{a}^2 - 2 V(a) = 0$ yields 
\be\label{GB_pot} 
	V(a) = \frac{a^{d+1}}{24 \lambda(M_+ - M_-)}\left[g_+(3+2\lambda g_+)^2 - g_-(3+2\lambda g_-)^2 \right] + \frac{\sigma}{2} 
\ee
for the thermalon potential.
Making use of \eqref{GB_char_poly}, we can put $V$ into a more useful form by reducing the order in $g_\pm$.  The result is,
\be 
V(a) = \frac{a^{d+1}}{24\lambda (M_+-M_-)} \left[(1+8\lambda\Lambda) g + (2+\lambda g) \frac{4 M}{a^{d-1}} \right] \Bigg|^+_- + \frac{\sigma}{2} \, .
\ee
Note that in the above, the factor $1+8\lambda \Lambda$ can be written as
\be 
1+8\lambda \Lambda = \lambda^2 \left( \Lambda_-^{\textrm{eff}} - \Lambda_+^{\textrm{eff}} \right)^2
\ee
where the term in parentheses could be interpreted as proportional to the difference between ``effective pressures" inside and outside the bubble.  Working further with this potential, we can obtain expressions for its $a$ derivatives, the first of which reads
\be\label{GB_pot_der} 
V'(a) = \frac{a^d}{24 \lambda (M_+-M_-)} \left[(d+1)(1+8\lambda\Lambda) g - \left[d-17 +2\lambda(d-5)g \right] \frac{M}{a^{d-1}} \right] \Bigg|_-^+ \, .
\ee
In obtaining \eqref{GB_pot_der} we have utilized the characteristic polynomial and its derivative to remove expressions involving $g_\pm'(a)$.  Expressions for higher derivatives of the potential can be obtained in the same manner, but for our analysis we shall need only the potential and its first derivative, since solving for consistent thermalon configurations amounts to the static condition $V(a_\star) = V'(a_\star)=0$.

\subsection{Stability}

Before moving on to an analysis of the thermodynamics of these systems, we pause to comment on the stability of thermalon configurations in GB gravity.  As was outlined in \cite{Camanho:2013uda}, there are two primary stability concerns: the dynamical instability of the thermalon solution and the possibility of the bubble escaping to infinity.  By expanding the junction condition about the thermalon solution $a=a_\star$,  to leading order it takes the form (cf. eq. (4.32) from \cite{Camanho:2013uda}),
\be 
\frac{\dot{a}^2}{2} +\frac{1}{2} k (a-a_\star)^2 = 0
\ee
where $k$  is given by
\be 
k = \frac{a_\star^2}{2} \left(\frac{\partial \tilde{\Pi}}{\partial H} \right) \frac{\partial ^2 \tilde{\Pi} }{\partial a^2} \Bigg|_{a = a_\star} \, .
\ee
and can be thought of as an effective Hooke's constant.
The sign of $k$ determines the stability of the thermalon configuration -- if it is positive, the thermalon is stable (the bubble can oscillate about $a=a_\star$ or remain fixed there) while a negative value of $k$ indicates that the thermalon is unstable (the bubble can expand, causing a phase transition).\footnote{The possibility of collapsing bubbles was discussed at length in \cite{Camanho:2013uda}.} 
Since we are interested in thermalon mediated phase transitions, we are interested only in cases where $k<0$.  In the case at hand, $k$ is proportional to the second derivative of the thermalon potential $V$. The particular expression for $k$ (or, equivalently, $V''$) is quite messy, but we have confirmed numerically that $k<0$ here, provided $\lambda >0$, for all physically relevant values of $\Lambda$.

 The second condition we are interested in is the possibility of the bubble escaping to infinity.  To this end, it is of interest to see how the speed of the bubble ($\dot{a}$) behaves in the limit of large $a$.  A consistent solution of the junction conditions in this limit yields,
\be 
H \approx \frac{a^{d-1}}{2(M_+-M_-)} \int_{\Lambda_-^{\textrm{eff}}}^{\Lambda_+^{\textrm{eff}}}  dx \, \Upsilon[x] \, .
\ee
where $H = (\sigma + \dot{a}^2)/a^2$ is related to the velocity of the bubble and approaches infinity as the bubble expands to infinity (for more details see the discussion around eq. (4.35) in \cite{Camanho:2013uda} and eq.~\eqref{pitildeexpr} above).  In the case considered here this expression has the simple form,
\be 
H \approx \frac{a^{d-1}}{2(M_+-M_-)} \frac{(1+8\lambda\Lambda)^{\frac{3}{2}}}{6 \lambda^2} = \frac{  a^{d-1} \lambda }{12   (M_+-M_-)}  \left( \Lambda_-^{\textrm{eff}} - \Lambda_+^{\textrm{eff}} \right)^3 \, .
\ee
 The limit $H\to \infty$ is consistent with the bubble expanding to infinity ($a\to \infty$) provided that we have $M_+ > M_-$, $\lambda >0$ and $\Lambda > -1/(8\lambda)$.  Looking at the last term we see that $H \sim  \Delta P_{\textrm{eff}}^3$.  This result fits well with our thermodynamic intuition: the bubble will be able to expand to infinity provided there is a positive difference between the effective pressures inside and outside of the bubble, which is the situation here.

\subsection{Thermodynamic picture}

In this section we wish to develop the extended phase space thermodynamics of the system under study.  In particular, we are interested in the case where the outer solution describes an asymptotically AdS spacetime, while the inner solution is asymptotically dS.  First, we consider the location of the bubble relative to the black hole horizon and the de Sitter horizon.  Recall that the thermalon corresponds to the static solution $V(a_\star) = V'(a_\star) = 0$.  We can then solve eqs. \eqref{GB_pot} and \eqref{GB_pot_der} to obtain $M_\pm$ as functions of $g_\pm$ and $a_\star$.  By substituting these results into the characteristic polynomial \eqref{GB_char_poly} we arrive at a system of two quadratic equations which can be solved for $g_\pm(a_\star)$.  The solution can be obtained analytically, but the expressions themselves are not worth quoting; however, it is worth mentioning  criteria used in determining which root is correct.  First, we cannot have $g_+(a_\star) = g_-(a_\star)$, since in this case we would not be  considering jump metrics and so no phase transition would occur.  Second, $g_+(a_\star)$ must be strictly negative, so that the metric function $f_+(r)$ is positive and properly describes an AdS spacetime outside the bubble. 
\begin{figure}[htp]
\centering
\includegraphics[scale=0.5]{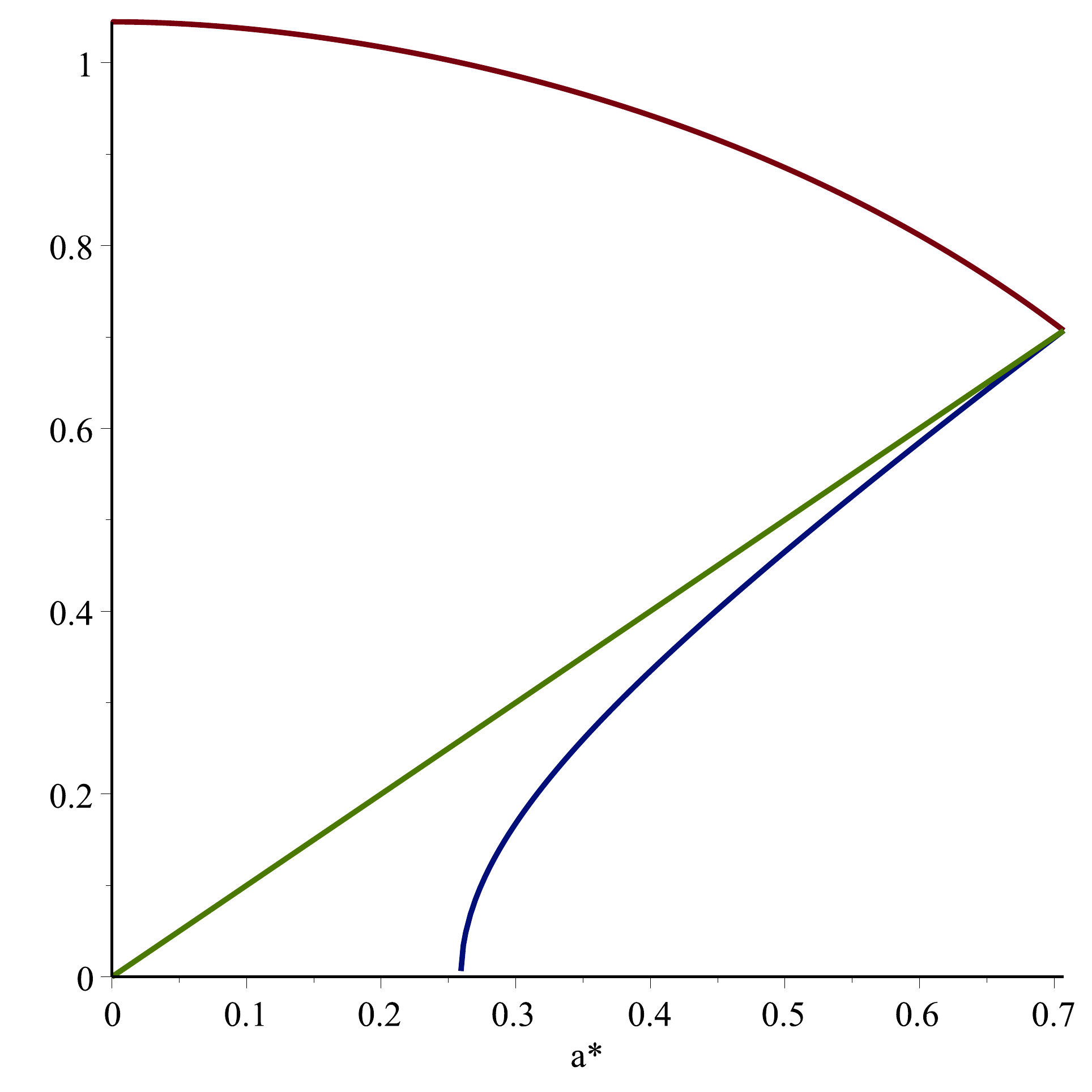}
\caption{A plot of $r_h$ (blue), $r_c$ (red) and $a_\star$ (green) as functions of $a_\star$ for $d=5$, $\lambda=0.1$, $\Lambda = 0.5$, $\sigma=1$.  We see that the bubble location, $a_\star$, is always found between the event horizon and the cosmological horizon until all three meet at the Nariai bound, $a_\star = 1/\sqrt{4 \Lambda}$.  The plot is qualitatively the same for $d > 5$. }
\label{fig:RhRc-astar}
\end{figure} 

Having solved for $g_\pm(a_\star)$, an explicit expression for $M_-(a_\star, \Lambda)$ can be obtained via the characteristic polynomial \eqref{GB_char_poly}.  We can then compare $a_\star$ with the event and cosmological horizon radii by solving $f_-(r_h) = f_-(r_c) = 0$ for $r_h$ and $r_c$ as functions of $M_-(a_\star,\Lambda)$.  We find that for all well-defined parameters the thermalon radius is larger than the event horizon radius, but smaller than the radius of the cosmological horizon up until the Nariai limit, at which all three occur at the same value.  The result is highlighted for the specific case $d=5$, $\Lambda = 0.5$, $\lambda=0.1$ and $\sigma = 1$ in Figure~\ref{fig:RhRc-astar}.   One consequence of this is that the cosmological horizon is not part of the spacetime, since outside the bubble the solution is given by the AdS branch.  We have, therefore, a Killing field which is timelike everywhere outside the event horizon allowing us to form a well-defined thermodynamic picture, free of the usual issues that  plague  dS spacetimes.  

We now focus on a development of the extended first law and Smarr formulae for this set up.  Since $\lambda$ here is dimensionful, we must consider it as a thermodynamic variable in the first law with a conjugate potential $\Psi$. Recalling that the pressure is given by eq. \eqref{pressure_def} and using this, along with the various properties of the inner black hole solution, it is straightforward to show that the extended first law \cite{Dolan:2013ft, Kastor:2010gq}
\be 
dM_- = T_- dS + V_- dP + \Psi_- d\lambda \, ,
\ee
is satisfied provided we identify 
\be 
V_- = r_h^{d-1}
\ee
as the thermodynamic volume\footnote{ The absence of a prefactor of the form $2 \pi^2$ follows from the conventions employed throughout this work.} and
\be
\Psi_- = \sigma r_h^{d-5} \left( \sigma - \frac{8 \pi r_h T_-}{d-4} \right)
\ee
for the potential conjugate to $\lambda$.  Note that here the Hawking temperature of the black hole is obtained in the standard way by requiring the absence of conical singularities in the Euclidean section ($t \to - i t_\mathrm{E}$),
\be 
T_- = \frac{f_-'(r)}{4 \pi} \Bigg|_{r=r_h}
\ee
while the entropy in this normalization is given by \cite{Camanho:2013uda}
\be\label{GB_entropy} 
S = 4 \pi r_h^{d-2} \left(\frac{1}{d-2} +   \frac{2 \sigma \lambda}{r_h^2 (d-4)} \right) \, .
\ee
These thermodynamic quantities satisfy the Smarr relation for the black hole 
\be 
(d-3)M_- = (d-2)T_-S - 2 V_- P + 2  \Psi_- \lambda 
\ee
as derived from scaling.

We now wish to develop the first law and Smarr relation for the quantities outside the bubble.  The outer first law and Smarr formula are given by
\ba\label{outerflawandsmarr}
dM_+ &=& T_+ dS + V_+ dP + \Psi_+ d\lambda \, , 
\nn\\
(d-3)M_+ &=& (d-2)T_+S - 2 V_+ P + 2  \Psi_+ \lambda \,,
\ea
which the thermodynamic quantities must satisfy.  As mentioned earlier, when the conditions for the existence of the thermalon are enforced (i.e. $V(a_
\star) = V'(a_\star) = 0$), we obtain expressions for $M_+$ and $M_-$ in terms of $a_
\star$ and $\Lambda$, the latter of which equivalently means we obtain an implicit relationship between $r_h$, $a_
\star$, and $\Lambda$:
\be\label{mmrelation} 
M_-(a_\star, \Lambda) = r_h^{d-1}\Upsilon\left[\frac{\sigma}{r_h^2} \right] \, .
\ee
This relationship gives us the freedom to write down thermodynamic expressions in terms of either $a_\star$ or $r_h$.  In the outer first law it is be easier to work directly with the former, since we have explicitly $M_+(a_\star, \Lambda)$. 
For the temperature, from the matching condition at the bubble we know that
\be\label{Tmatch}
T_+ = \sqrt{\frac{f_+(a_\star)}{f_-(a_\star)}} T_- \, .
\ee
The entropy that appears in the outer first law is simply that given in \eqref{GB_entropy} for the black hole, since the bubble does not contribute to the entropy.  

It is only practical to compute the expressions of the quantities in \eqref{outerflawandsmarr}
analytically in $d=5$, but even then they are particularly messy---especially $V_+$ and $\Psi_+$.  For these reasons we have verified a consistent solution of \eqref{outerflawandsmarr} numerically. In general, $V_+$ is not independent of $P$, which is (at least in part) due to the relationship  \eqref{mmrelation} including $P$ as the zeroth order contribution in $\Upsilon\left[\sigma/r_h^2 \right]$.

\subsection{Criticality \& phase phenomena}

With the tools developed in the previous sections we are now situated to perform the extended phase space analysis for these transitions.  We focus on the case $\sigma=1$ and $d=5$.  The thermodynamically preferred state at a given temperature and pressure is that which minimizes the Gibbs free energy. In \cite{Camanho:2013uda} the Euclidean action for the thermalon configuration was shown to be
\be 
{\cal I} = \beta_+ M_+ - S \,.
\ee
In general, the Euclidean action of the thermalon configuration is divergent; however, it can be suitably regularized by subtracting the (infinite) contribution of thermal AdS space   yielding the result above. It is then the case that the Gibbs free energy is given by
\be 
G = M_+ - T_+S \,.
\ee

Here we will be comparing the free energy of the thermalon configuration to that of pure AdS space, the latter being identically zero due to the fact that it was used in the background subtraction.  Despite the naive impression that there are six independent parameters ($M_\pm, \beta_\pm, a_\star$ and $P$), there are in fact only two, $T_+$ and $P$, since there are four equations relating the six quantities: $V(a_\star)=V'(a_\star)=0$, the Hawking condition for the inner black hole, and the matching of thermal circles \eqref{Tmatch} imposed at the junction.  In general it is difficult or impossible to write $M_+$ and $S$ as explicit functions of $T_+$ and $P$, therefore we studied the behaviour of $G$ numerically.

\subsubsection{Negative pressure: thermal AdS to dS  black hole transitions}

We begin by considering the situation in which  the thermalon separates spacetime into regions with AdS asymptotics outside and dS asymptotics inside.  We build upon the approach of \cite{Camanho:2015zqa} by performing an exhaustive analysis of the pressure parameter space.  
\begin{figure}[htp]
\centering
\includegraphics[width=0.7\textwidth]{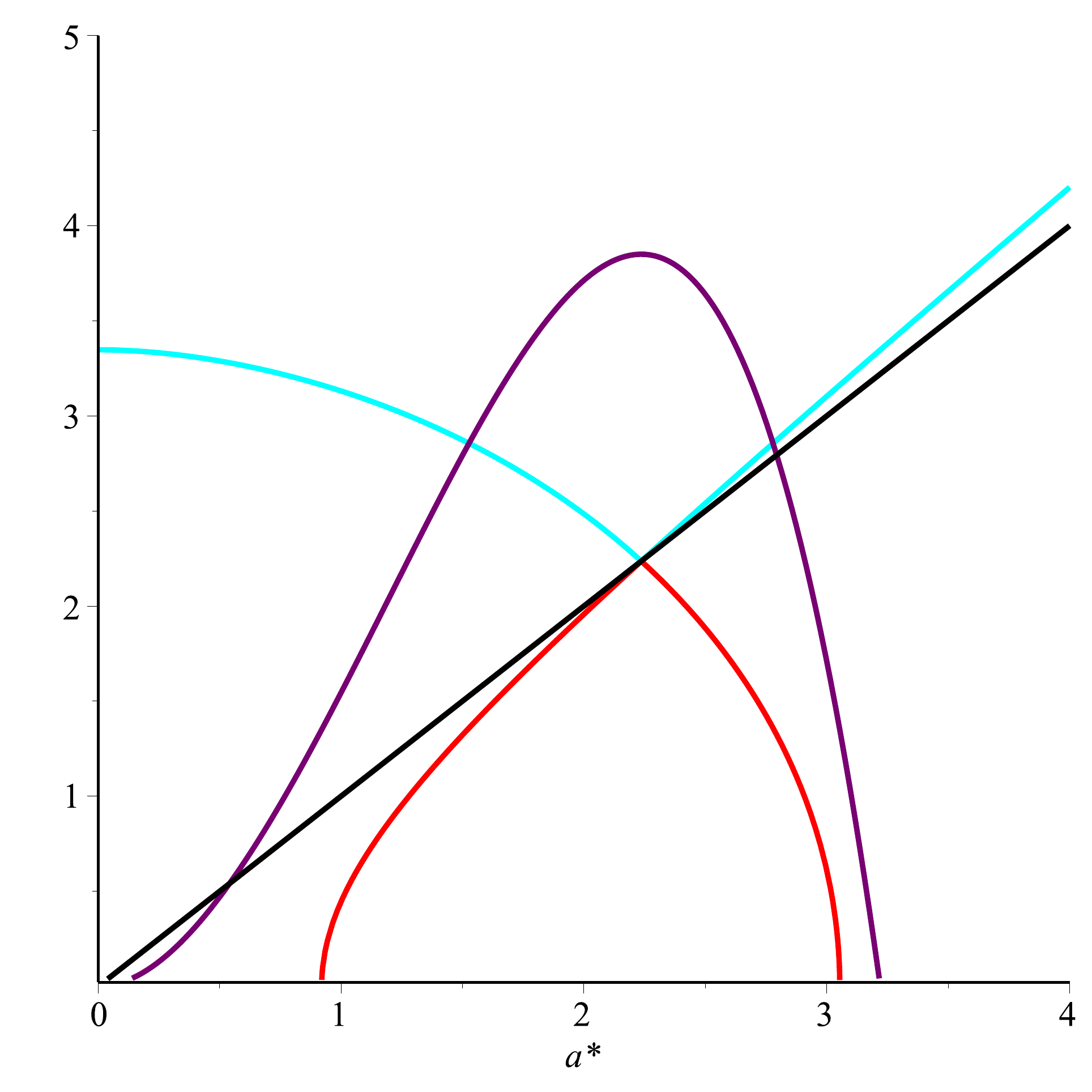}
\caption{A plot displaying $a_\star$ (black), the cosmological horizon (cyan), the event horizon (red) and the mass parameter $M_-$ (purple) as functions of $a_\star$,  in units of the Planck length. The Nariai limit corresponds to the point where the black, cyan, and red curves meet.  This plot corresponds to $P=-0.1$ and $\lambda=1.35$; plots for other parameter values are qualitatively similar.  }
\label{alocal}
\end{figure}
 
We first consider the parameter range over which we can obtain a sensible solution of the four equations governing the thermalon. A representative plot is shown in Figure~\ref{alocal} which highlights the salient features.  We see that for some range of $a_\star$ there is a consistent solution where the inner de Sitter space has a cosmological horizon, an event horizon and the mass parameter $M_-$ is positive.  Outside of this region (values of $a_\star$ that lie beyond the edges of the red curve) there is no consistent solution and bubbles of these sizes cannot form. To the right of the Nariai limit, the conditions $V(a_\star)=V'(a_\star)=0$ are satisfied. However in this region $\Pi^+ = - \Pi^-$, and so the junction conditions cannot be satisfied without the addition of a shell of stress-energy.  Furthermore, it can be shown that $\Pi^+<0$ for $a_\star>a_{\textrm{Nariai} }$, since the only zero of $\Pi^+$ occurs at the Nariai limit and  its slope as a function of $a$ is negative there.  Consequently any such shell must be composed of exotic matter, since $\rho \sim \Pi^+ - \Pi^- = 2\Pi^+ <0$.

\begin{figure}[htp]
\centering
\includegraphics[width=0.45\textwidth]{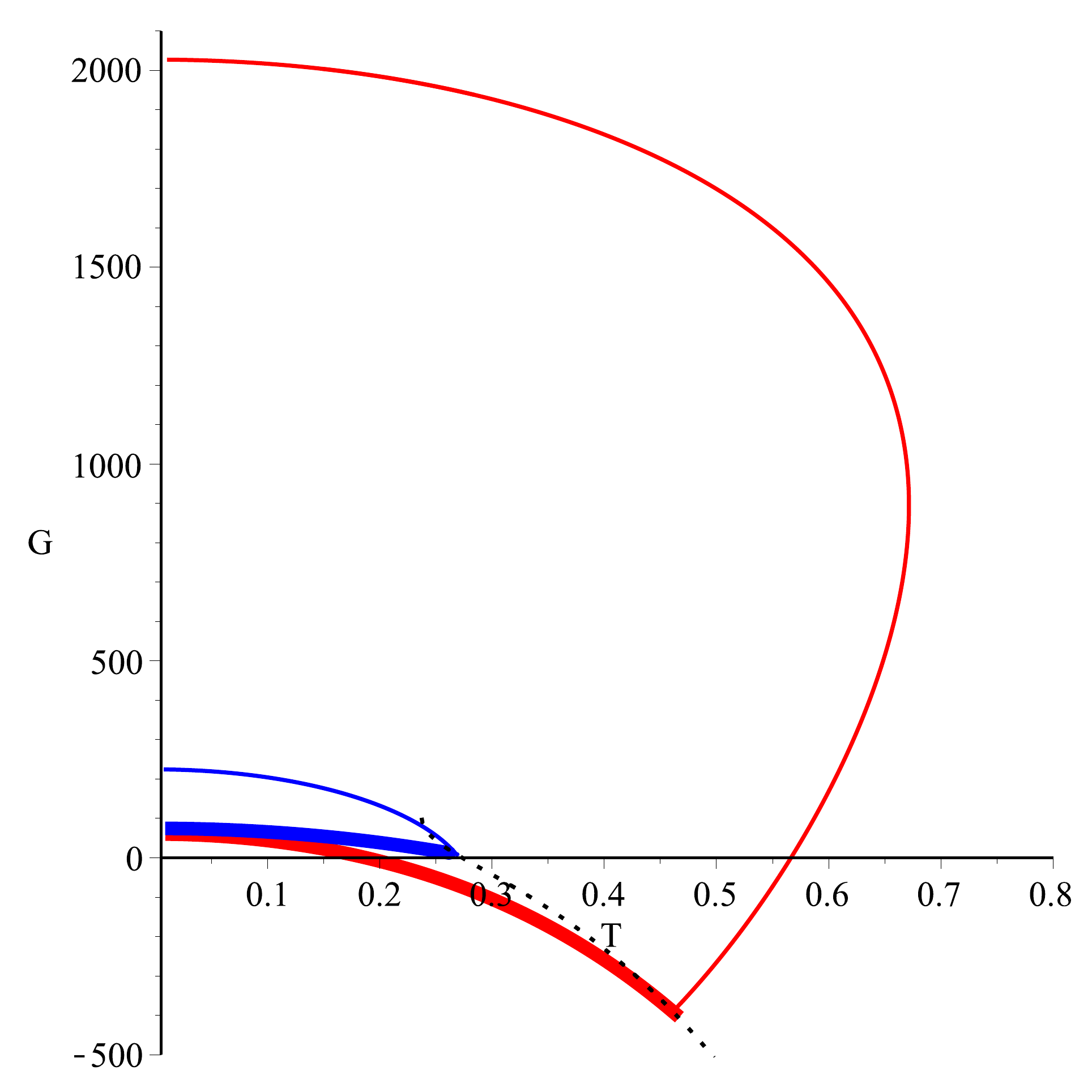}
\includegraphics[width=0.45\textwidth]{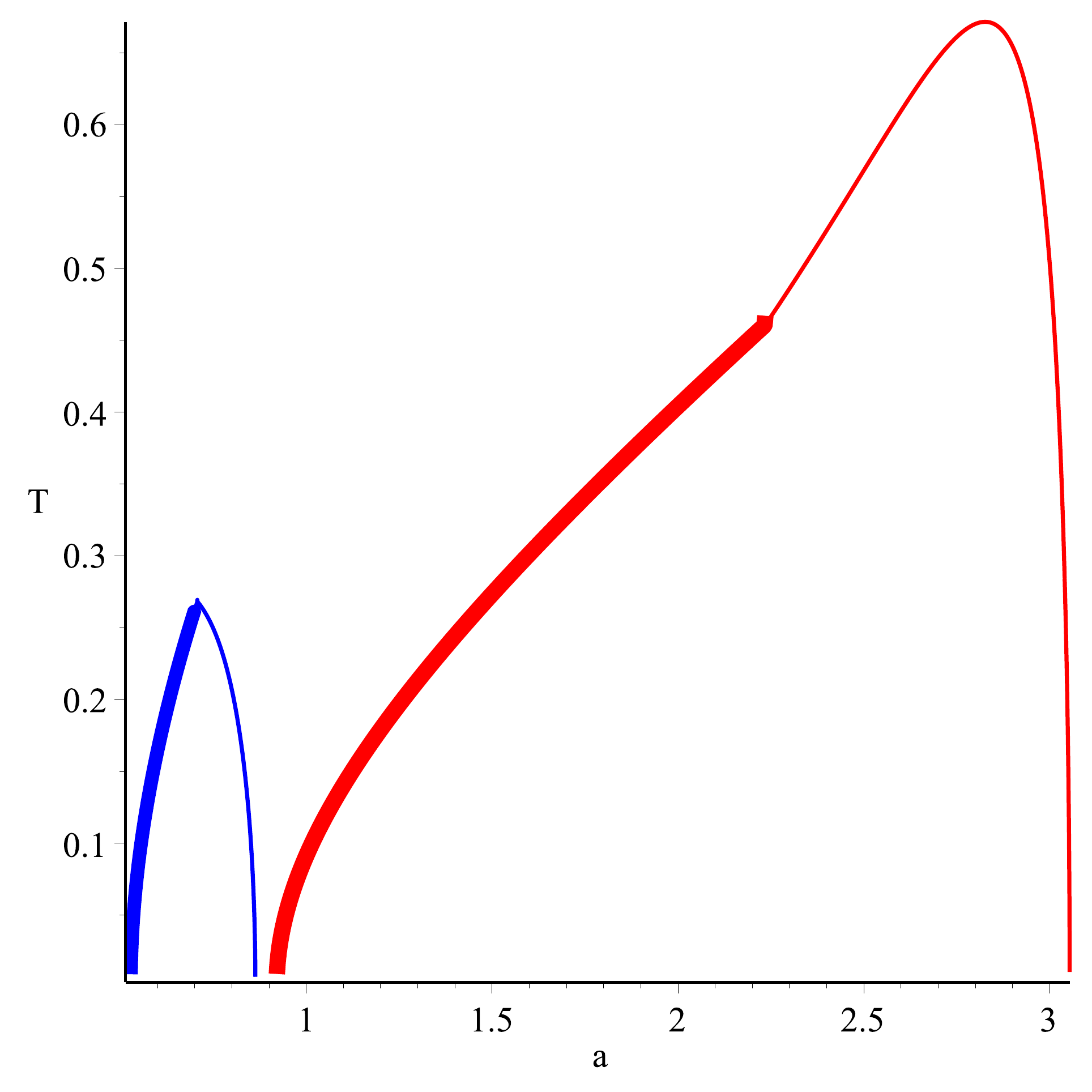}
\includegraphics[width=0.45\textwidth]{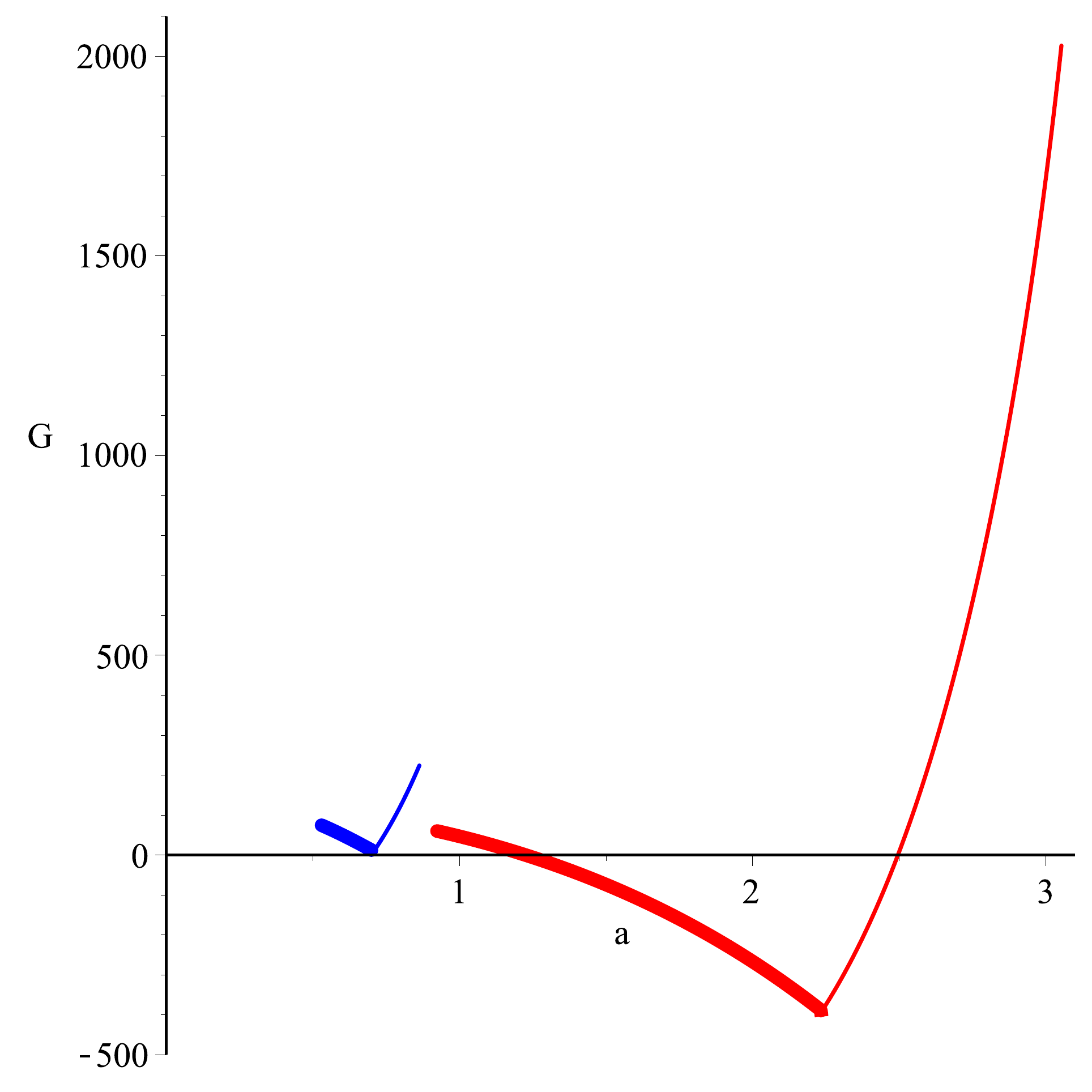}
\caption{{\bf AdS to dS transition: pressure effects: $\lambda = 1.35$}.  The red curves correspond to $P=-0.1$ while the blue curves correspond to $P=-1$.  {\it Upper left}: A plot of the free energy vs. $T_+$. For $P=-0.1$ a thermalon mediated phase transition is possible over a range of temperature, while it is not possible for $P=-1$.  In each case the thin upper branch is unphysical, corresponding to $\Pi^+ = - \Pi^-$.  The dotted black line corresponds to the Gibbs free energy for the Nariai limit as a function of pressure.  {\it Upper right}: A plot of the temperature, $T_+$ vs. $a_\star$ n both the blue and red curve, the cusp corresponds to the Nariai limit.  {\it Bottom}: A plot showing the Gibbs free energy as a function of $a_\star$ with the cusps again corresponding to the Nariai limit.  All quantities are measured in units of the Planck length.   }
\label{pressure_effects}
\end{figure}

The behaviour of the Gibbs free energy is very interesting and is influenced by both the pressure and the Gauss-Bonnet coupling.  First let us consider the effect the pressure has on the Gibbs free energy.  Figure~\ref{pressure_effects} shows free energy and temperature plots for $\lambda=1.35$ with the red and blue curves corresponding to $P=-0.1$ and $P=-1$, respectively.  These plots illustrate a general feature: for a given fixed $\lambda$, thermalon mediated phase transitions are possible for values of $P$ near zero, while for $P$  increasingly negative there is a point after which the free energy is strictly positive, and no phase transitions can occur---in Figure~\ref{pressure_effects} this happens for pressures near $P=-1$. As the pressure becomes closer to zero, the range of temperatures over which the thermalon mediated phase transitions can occur becomes larger.  This suggests the possibility of observing AdS $\to$ Minkowski space phase transitions in the limit where $P=0$, which we explore in the following section.

\begin{figure}[htp]
\centering
\includegraphics[width=0.45\textwidth]{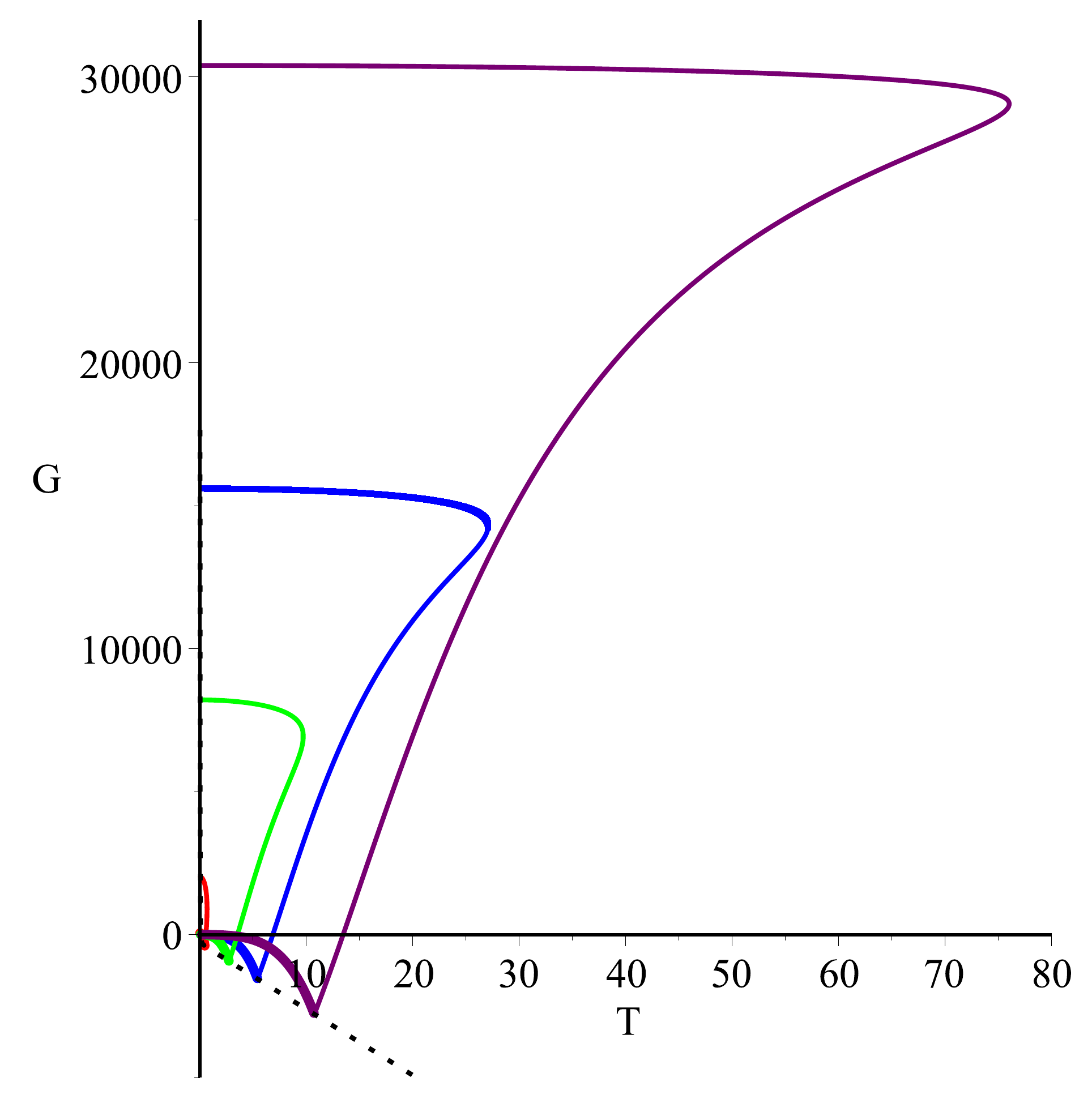}
\includegraphics[width=0.45\textwidth]{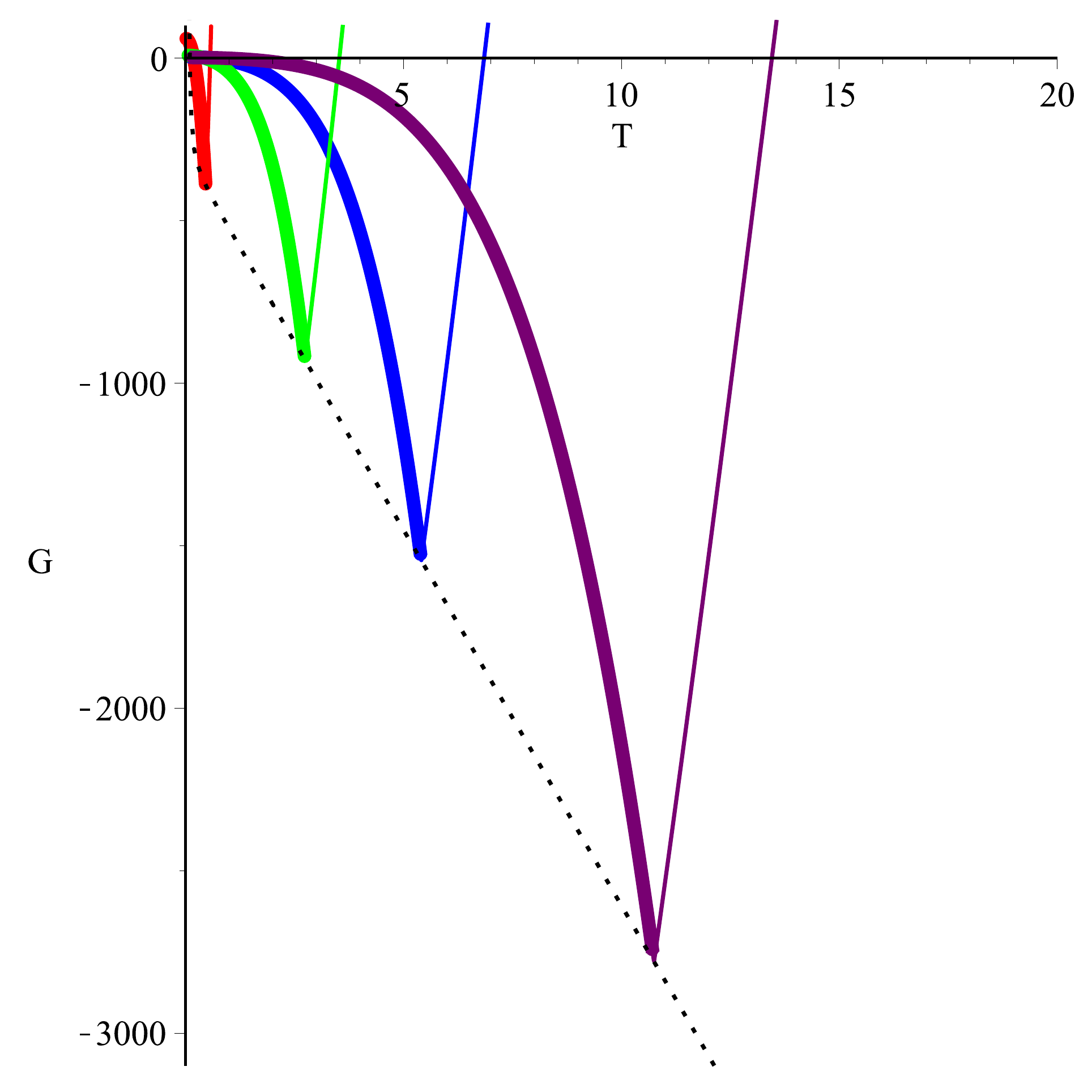}
\caption{{\bf AdS to dS transition: $\lambda$ effects}: $P=-0.1$.  The above plots show the free energy vs temperature ($T_+$) for $\lambda = 0.05, 0.1, 0.2, 1.35$ (from right to left) with the right plot being just a zoomed-in version of the left.  Thermalon mediated phase transitions are possible over a wider range of temperatures for larger values of $\lambda$.  The physical parts of the curves are the thick ones to the left of the cusps.
The dotted black line corresponds to the Nariai limit as a function of $\lambda$.   Quantities are measured in units of the Planck length. }
\label{lambda-effects}
\end{figure}

From Figure~\ref{pressure_effects} it appears as through the free energy is double valued, with the possibility of small and large bubbles for each value of temperature.  However, this is not the case.  The large bubble branch turns out to be unphysical (or would require exotic matter): while it satisfies $V(a_\star)=V'(a_\star)=0$ it does not satisfy $\Pi^+ = \Pi^-$. The same is true for the branches to the right of the cusp in Figure~\ref{lambda-effects}. The cusp in the Gibbs free energy vs. temperature curve corresponds to parameter values that yield the Nariai limit.

The next feature we examine is how the free energy depends on $\lambda$, with representative results shown in Figure~\ref{lambda-effects}.  From these plots we see that for various ranges of temperatures the free energy is negative, indicating the possibility of thermalon mediated phase transitions.  We note that the range of temperatures over which these transitions are possible increases as $\lambda$ is made smaller.
 \begin{figure}[htp]
\centering
\includegraphics[width=0.6\textwidth]{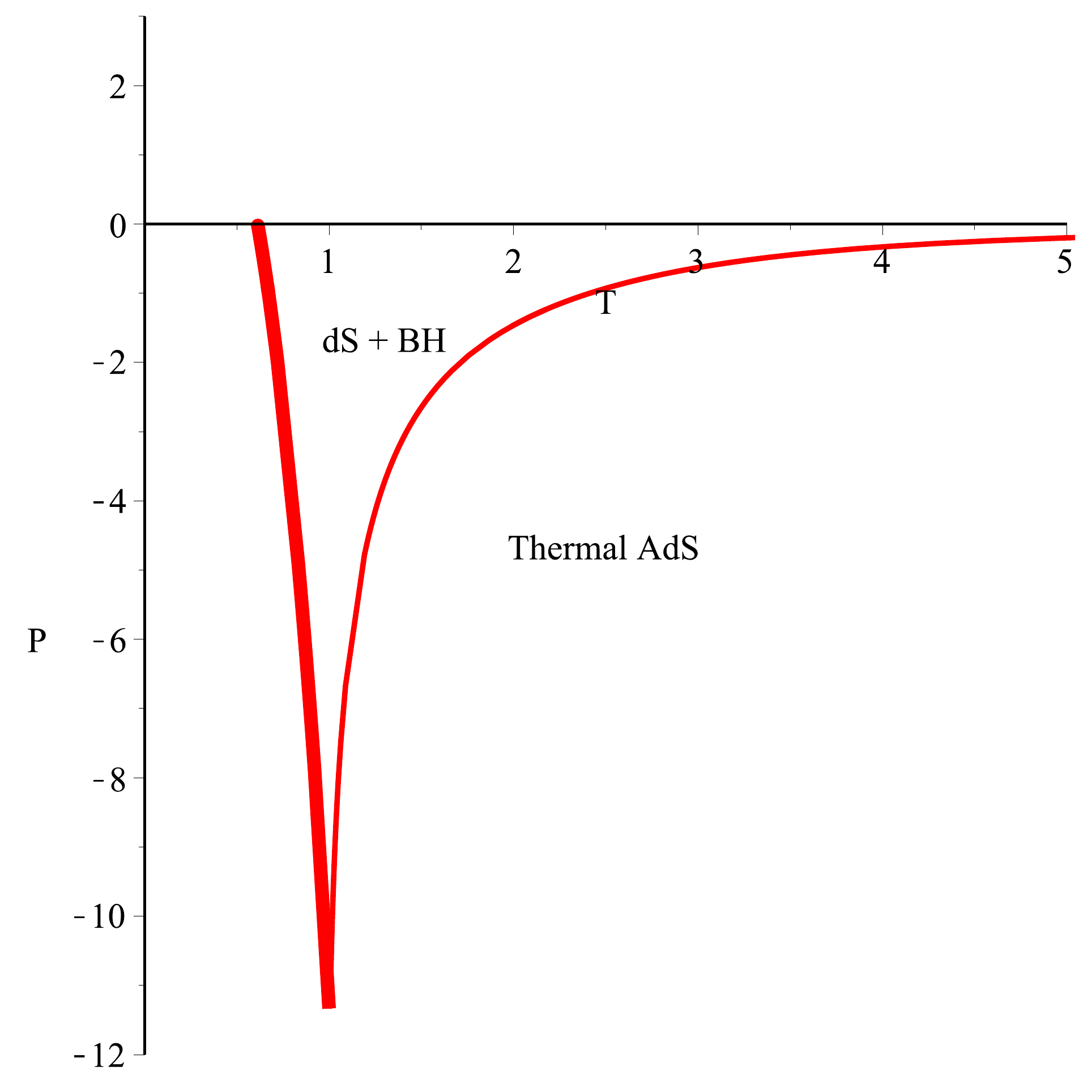}
\caption{{\bf AdS to dS transition: $P-T$ plane}: $\lambda = 0.1$. For parameter values inside the red curve a thermalon mediated phase transition is permitted, while parameters outside of this wedge correspond to thermal AdS space---no phase transition is possible.  The cusp corresponds to the Nariai limit; the physical curve is the thick one at the right.} 
\label{coexistence_plot}
\end{figure}

We can attain further insight by considering these phase transitions in the $P-T$-plane, as shown in Figure~\ref{coexistence_plot} for $\lambda=0.1$.  Here, the red curve marks the parameter values for which the free energy of the thermalon is identically zero.  Within the region bounded by the left-most part of the red curve and the Nariai temperature (the cusp of the red curve), the free energy of the thermalon is negative and a phase transition can occur.   Outside of this region, either the free energy of the thermalon is positive or no physical thermalon solution exists, and thus the thermal AdS space will not undergo a thermalon mediated phase transition.   The piece of the red curve to the right of the cusp corresponds to the zeros of the unphysical (or exotic matter) branch.

 We pause here to make a cautionary remark.  While Figure~\ref{coexistence_plot} is similar to coexistence plots,   it is important to distinguish these thermalon mediated phase transitions from the type of phase transitions we normally study using the tools of extended phase space thermodynamics.  Typically one compares a number of configurations all of which are in thermal equilibrium.  The key difference for  the thermalon is that it is unstable---once it forms it rapidly expands to infinity, changing the asymptotics of the spacetime.  Furthermore, there is no regular thermalon solution with dS asymptotics outside and AdS asymptotics inside.  In other words, this phase transition can only proceed in the direction of thermal AdS to a de Sitter black hole; the reverse process is not possible.  The consequence is this: if one wishes to read Figure~\ref{coexistence_plot} in a manner similar to how a coexistence plot would be read, it must be kept in mind that the only physical interpretations correspond to adjusting parameters so that the state of the system \emph{enters} the region bounded by the red curve, and never exits it.  In other words, it would be incorrect to say that the plot physically describes a $AdS \to dS + BH \to AdS$ re-entrant phase transition for a single spacetime.  Rather, in the context of an ensemble of spacetimes,  
 as temperature is monotonically increased, we go from stable thermal AdS, to unstable AdS in which a black hole forms in a de Sitter environment, and then back to stable thermal AdS.   

\begin{figure}[htp]
\centering
\includegraphics[width=0.6\textwidth]{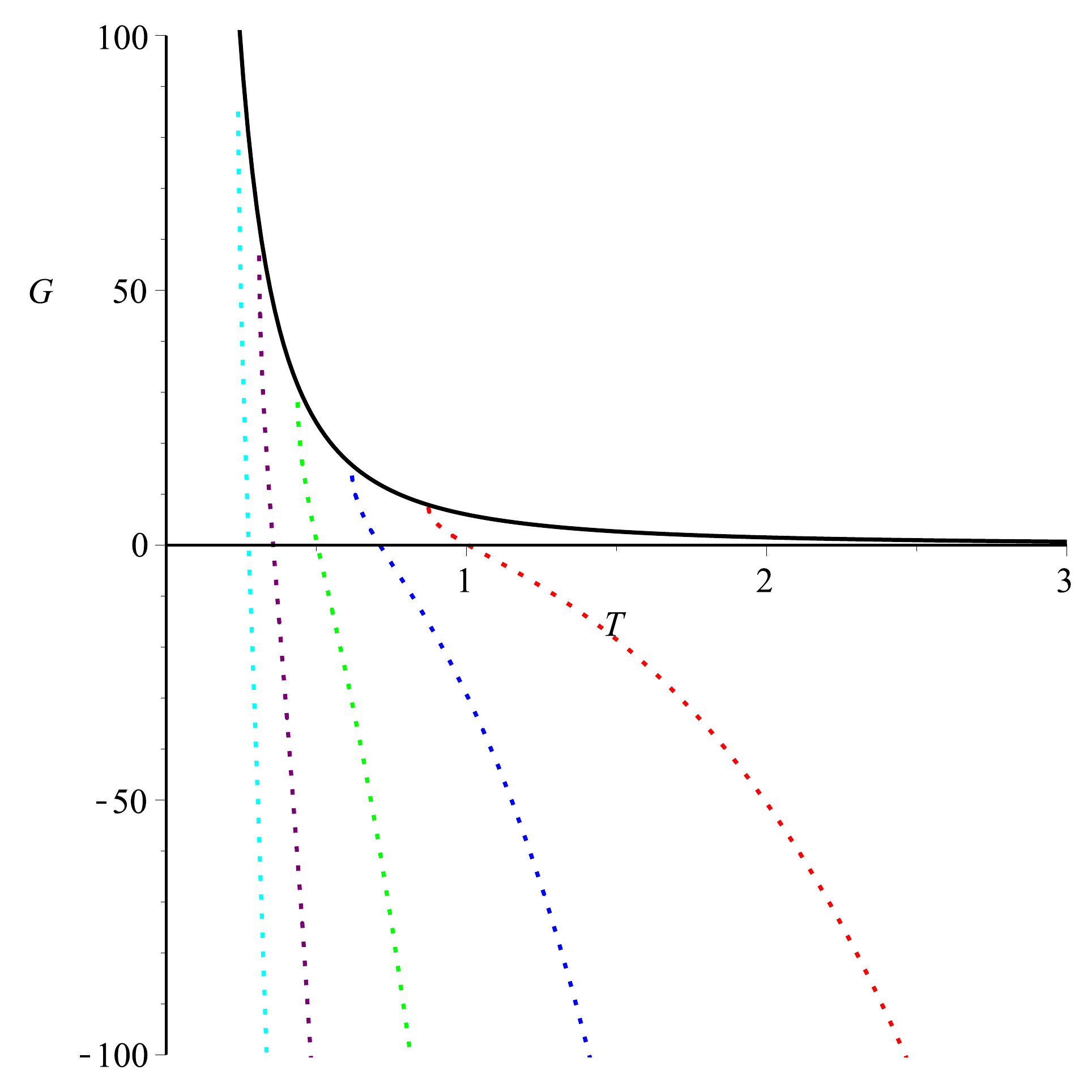}
\caption{{\bf AdS to dS transition: Nariai Gibbs free energy}. The dotted lines display the Gibbs free energy at the Nariai limit for $\lambda=0.1, 0.2, 0.4, 0.8, 1.35$ (right to left, respectively).  The solid black curve displays the locus of points corresponding the limit $P \to -\infty$ of the Nariai Gibbs free energy.  The quantities are measured in units of the Planck length. } 
\label{nariai_plot}
\end{figure}
  
Although the discussion above focused on the specific case $\lambda=0.1$, the ideas are quite general, and results are qualitatively identical for all $\lambda > 0$, as we shall now discuss.  For example, one universal feature is that, for all $\lambda > 0$, there is a pressure beyond which no phase transition will occur or alternatively, for  which   the Gibbs free energy is always positive.  To see this, we can study the Gibbs energy in the Nariai limit which we denote $\tilde{G} = \tilde{M}_+ - \tilde{T}_+\tilde{S}$, where the tilde represents the quantities are evaluated in this extremal limit.   This will be helpful since, as we saw in the earlier discussion, the (physical) Gibbs free energy terminates at the Nariai limit, and this point corresponds to the minimum of  the Gibbs free energy. Note that, even though $T_-$ is zero in this limit, $T_+$ remains finite since 
\be 
\tilde{T}_+ = \sqrt{\frac{f_+(r_h)}{f_-(r_h)}}T_-
\ee
and $T_-$ and $\sqrt{f_-(r_h)}$ approach zero at the same rate. The expression for $\tilde{G}$ takes the form,
\be 
\tilde{G} = \frac{192 P^2 \lambda^2 + 9 - 84P\lambda - (4-48\lambda P)\sqrt{9-24\lambda P}}{48 \lambda P^2} 
\ee
and 
\be 
\tilde{T}_+ = \frac{1}{8\pi\lambda} \sqrt{\frac{48\lambda P-18}{P}} \,.
\ee
 For a given fixed value of $\lambda$, as $P \to 0_-$, we have  $\tilde{T}_+ \to \infty$ and $\tilde{G} \to  -\infty$, so the Gibbs free energy will always be negative for negative pressures sufficiently close to zero.  On the other hand, as $P \to - \infty$ we have
\be 
\tilde{T}_+ \to \frac{1}{2 \pi} \sqrt{\frac{3}{\lambda}}\, , \quad \tilde{G} = 4\lambda + {\cal O}\left(\sqrt{T}\right)
\ee
meaning the Gibbs energy at the Nariai limit will be positive for sufficiently large negative pressures.  
What we glean from these two cases is that, regardless of the value of $\lambda$ (so long as it is positive), the plot of $\tilde{G}(T)$ will always resemble the dotted lines of Figure~\ref{nariai_plot}: it will be positive for large negative pressures and negative for sufficiently small pressures. For the pressures that satisfy $\tilde{G} >0$, there will be no phase transitions.  In other words, for any given $\lambda$ there will exist a minimum pressure $P_0$ such that for all $ P < P_0$ the free energy of the thermalon is always positive, and no phase transitions take place.

\subsubsection{Vanishing pressure: thermal AdS to asymptotically flat black hole transitions}

\begin{figure}[htp]
\centering
\includegraphics[scale=0.33]{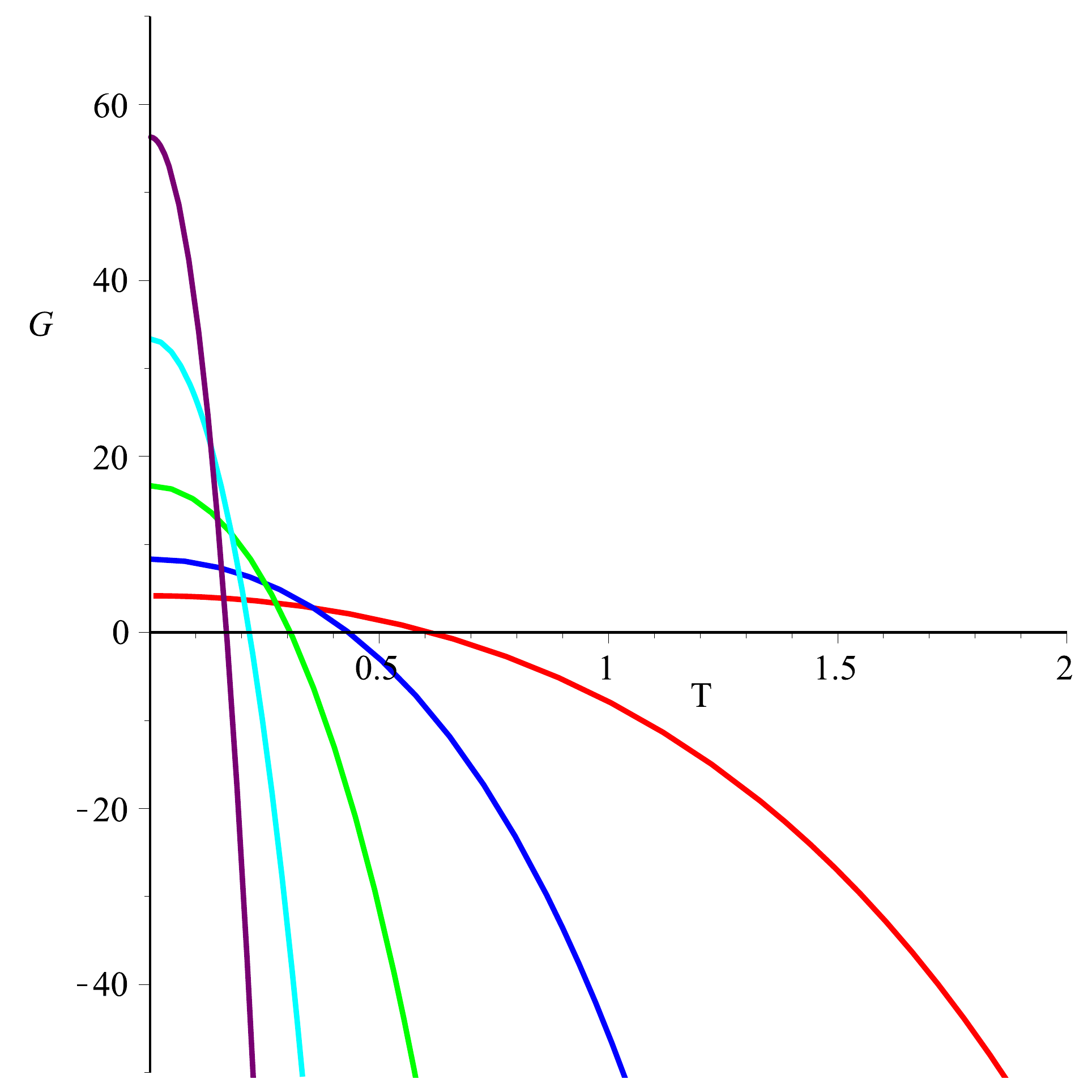}
\includegraphics[scale=0.33]{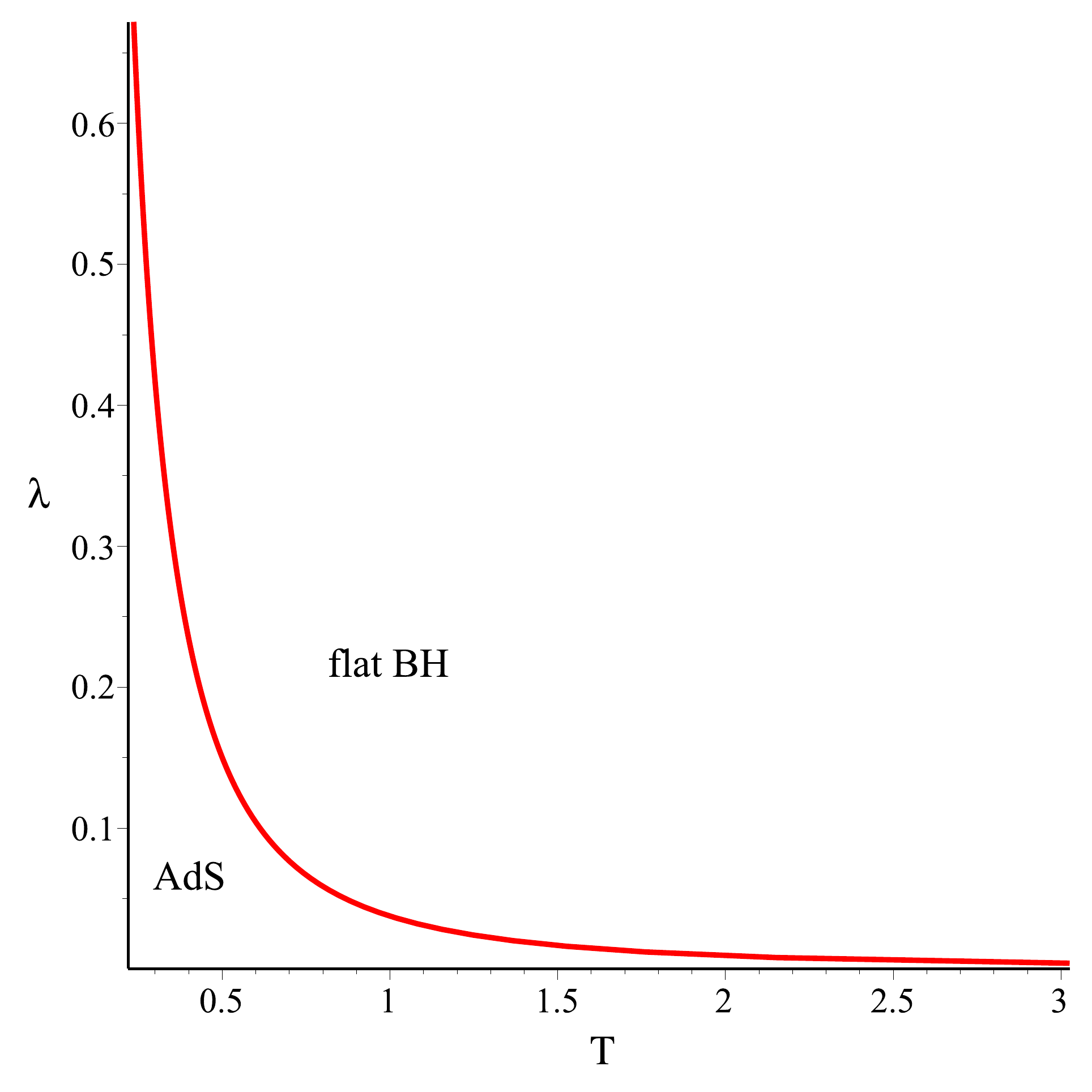}
\caption{{\bf AdS to flat space transition}: {\it Left}: $G$ vs. $T$ plot for $P=0$ showing $\lambda=0.1, 0.2, 0.4, 0.8, 1.35$ (bottom to top in $y$-intercept).  For each value of $\lambda$ there is a temperature above which the thermalon-mediated transition can occur. {\it Right}:  The coexistence plot in $\lambda-T$ space for the AdS to asymptotically flat black hole transition.  Below  the red line, the thermodynamically preferred state is thermal AdS space, while above the red line an asymptotically black hole is thermodynamically preferred.  The coexistence plot can only be read from left to right, and not from right to left, since the thermalon is dynamically unstable.}
\label{zeroPtrans}
\end{figure}

The analysis above hinted towards the possibility of observing transitions between thermal AdS space and an asymptotically flat black hole when $P=0$.  The junction conditions discussed earlier are in no way changed by specializing to the specific case $P=0$, and so we can proceed as before.  We note that, for $P=0$, the effective cosmological constants are
\be\label{effectivecosmo2} 
	\Lambda_\pm^{\textrm{eff}} = - \frac{1 \pm 1}{2 \lambda} 
\ee
or in other words  $\Lambda_+^{\textrm{eff}}=-1/\lambda$ and $\Lambda_-^{\textrm{eff}}=0$.
The Gibbs energy remains $G=M_+-T_+S$, and Figure~\ref{zeroPtrans} shows representative plots of $G$ vs. $T$ for various values of $\lambda$. 

Since the $``-"$ branch describes an asymptotically flat black hole, we do not face the complications associated with the Nariai limit here.  In Figure~\ref{zeroPtrans} this amounts to the fact that the Gibbs free energy does not terminate at a particular temperature, but is well-defined for all positive temperatures.  This feature is highlighted in the right plot of Figure~\ref{zeroPtrans}, which shows a $\lambda - T$ coexistence plot for $P=0$.  One must keep in mind that, because the thermalon is not an equilibrium configuration, the coexistence plot can only be read from left to right---an asymptotically flat black hole will not spontaneously decay to AdS space by this mechanism. As a consequence of this, we see that regardless the value of $\lambda >0$, there will always be a temperature above which it becomes thermodynamically favourable for the thermal AdS vacuum to decay to an asymptotically flat black hole.    

One might be concerned about this transition from the point of view of energy: thermal AdS space is decaying into a spacetime containing an asymptotically flat black hole.  However, it is important to keep in mind that the ``true" cosmological constant, $c_0$ in the characteristic polynomial, is zero here.  The asymptotically AdS structure of the outer branch is a result of the non-zero Gauss-Bonnet coupling, as observed in eq.~\eqref{effectivecosmo2}.  The equivalent transition could not occur in Einstein gravity.

\subsubsection{Positive pressures: thermal AdS to AdS black hole transitions}

Considering the stability constraints discussed earlier, there is a small range of positive pressures for which the thermalon mediated phase transitions can occur.  Specifically, our constraints were found earlier to be $\Lambda > -1/(8\lambda)$, which in terms of pressure reads $ P < 1/(4\lambda)$.  For pressures in the range $0 < P < 1/(4\lambda)$ both branches of the GB solution admit AdS asymptotics, and the thermalon then describes a transition between thermal AdS space and an AdS black hole, in some ways analogous to the Hawking-Page transition.  The behaviour of the free energy in this case is qualitatively identical to that shown in Figure~\ref{zeroPtrans}, and so we do not replicate the plot again here.  Due to the similarities, one may wonder whether there is some competition between the Hawking-Page and thermalon mechanisms. 

 As it turns out, such a situation does not arise. 
In the case of the Hawking-Page transition, one finds that an AdS black hole is thermodynamically favoured to thermal AdS space above a certain critical temperature.  In order to make sense of this transition, the thermal AdS space and AdS black hole should have the same asymptotic structure, i.e. the same (effective) cosmological constant.  In the case of the thermalon mediated phase transitions, one is again considering a transition between thermal AdS space and an AdS black hole; however the difference in this case is that the effective cosmological constants of the thermal AdS space and the AdS black hole \emph{are different}.  

This makes the comparison between the two transitions a questionable one for the following reasons.  Given values for $\Lambda$ and $\lambda$, these correspond to some $\Lambda^{\rm eff}_{\pm}$.  If we are interested in considering a thermalon mediated phase transition, then we consider the thermal AdS vacuum to have cosmological constant $\Lambda^{\rm eff}_{+}$, which then decays to a black hole with cosmological constant $\Lambda^{\rm eff}_{-}$.  However, this poses a problem for the Hawking-Page transition, since for thermal AdS space with $\Lambda$ and $\lambda$ corresponding to $\Lambda_{+}^{\rm eff}$, the theory does not permit a black hole solution which has the \emph{same} cosmological constant \emph{and} coupling constants $\Lambda$ and $\lambda$.  In other words, the branch of the Gauss-Bonnet solution that is asymptotically described by $\Lambda_{+}^{\rm eff}$ does not describe a black hole, and so the Hawking-Page transition would not occur for it.  In order to describe a Hawking-Page transition for thermal AdS space with cosmological constant $\Lambda_{+}^{\rm eff}$, one would have to consider a theory with different values of $\Lambda$ and $\lambda$ such that $\Lambda^{\rm eff}_{+}(\Lambda, \lambda) = \Lambda^{\rm eff}_{-}(\tilde{\Lambda}, \tilde{\lambda})$.       
To summarize, it does not seem sensible to talk about a comparison between the Hawking-Page and thermalon mediated phase transitions for a given theory specified by value of $\Lambda$ and $\lambda$.

\section{Conclusions}

We have performed an analysis of thermalon mediated phase transitions in extended thermodynamic phase space.  In addition to showing the results previously studied in \cite{Camanho:2013uda, Camanho:2015ysa,Camanho:2015zqa} are consistent with the extended phase space paradigm, we have found a number of new and interesting features of these transitions.  In terms of the effect of the thermodynamic pressure, we have shown that for any given value of the Gauss-Bonnet coupling, for large enough negative pressures (i.e. large, positive cosmological constants) the phase transitions are not possible.

In addition to considering thermalon phase transitions in the case where the inner solution is de Sitter, we have also considered the possibility where the inner solution is asymptotically flat.  Here we have found that thermal AdS space can undergo a thermalon mediated phase transition to an asymptotically flat black hole spacetime.  In contrast to the de Sitter case, where the phase transitions are only possible over a small range of temperatures, in the asymptotically flat case, a thermalon transition is possible at arbitrarily large temperatures.

\section*{Acknowledgments}
We thank Jos\'e Edelstein for bringing to our attention an issue regarding the interpretation of the unphysical branch of the Gibbs energy.  This work was supported in part by the Natural Sciences and Engineering Research Council of Canada.

\providecommand{\href}[2]{#2}\begingroup\raggedright\endgroup

\end{document}